\documentclass[prb,twocolumn,superscriptaddress,floatfix,letterpaper,aps]{revtex4-2}

\usepackage[utf8]{inputenc}
\usepackage{natbib}
\usepackage{amsmath}
\usepackage{amsbsy}
\usepackage{amssymb}
\usepackage{scrextend}
\usepackage{graphicx}
\usepackage{color}
\usepackage{multirow, array}
\usepackage{float}
\usepackage{hyperref}
   \hypersetup{colorlinks=true}

\begin{document}
\title{Band structure, superconductivity and polytypism in AuSn$_4$}
\author{Edwin Herrera}
\affiliation{Laboratorio de Bajas Temperaturas y Altos Campos Magn\'eticos, Departamento de F\'isica de la Materia Condensada, Instituto Nicol\'as Cabrera and Condensed Matter Physics Center (IFIMAC), Unidad Asociada UAM-CSIC, Universidad Aut\'onoma de Madrid, E-28049 Madrid,
Spain.}

\author{Beilun Wu}
\affiliation{Laboratorio de Bajas Temperaturas y Altos Campos Magn\'eticos, Departamento de F\'isica de la Materia Condensada, Instituto Nicol\'as Cabrera and Condensed Matter Physics Center (IFIMAC), Unidad Asociada UAM-CSIC, Universidad Aut\'onoma de Madrid, E-28049 Madrid,
Spain.}

\author{Evan O'Leary}
\affiliation{Ames Laboratory and Department of Physics $\&$ Astronomy, Iowa State University, Ames, IA 50011}
\author{Alberto M. Ruiz}
\affiliation{Instituto de Ciencia Molecular (ICMol), Universitat de Val\'encia, Catedr\'atico Jos\'e Beltr\'an 2, 46980 Paterna, Spain.}

\author{Miguel \'Agueda}
\affiliation{Laboratorio de Bajas Temperaturas y Altos Campos Magn\'eticos, Departamento de F\'isica de la Materia Condensada, Instituto Nicol\'as Cabrera and Condensed Matter Physics Center (IFIMAC), Unidad Asociada UAM-CSIC, Universidad Aut\'onoma de Madrid, E-28049 Madrid,
Spain.}

\author{Pablo Garc\'ia Talavera }
\affiliation{Laboratorio de Bajas Temperaturas y Altos Campos Magn\'eticos, Departamento de F\'isica de la Materia Condensada, Instituto Nicol\'as Cabrera and Condensed Matter Physics Center (IFIMAC), Unidad Asociada UAM-CSIC, Universidad Aut\'onoma de Madrid, E-28049 Madrid,
Spain.}

\author{V\'ictor Barrena}
\affiliation{Laboratorio de Bajas Temperaturas y Altos Campos Magn\'eticos, Departamento de F\'isica de la Materia Condensada, Instituto Nicol\'as Cabrera and Condensed Matter Physics Center (IFIMAC), Unidad Asociada UAM-CSIC, Universidad Aut\'onoma de Madrid, E-28049 Madrid,
Spain.}

\author{Jon Azpeitia}
\affiliation{Instituto de Ciencia de Materiales de Madrid, Consejo Superior de Investigaciones Cient\'{\i}ficas (ICMM-CSIC), Sor Juana In\'es de la Cruz 3, 28049 Madrid, Spain.}

\author{Carmen Munuera}
\affiliation{Instituto de Ciencia de Materiales de Madrid, Consejo Superior de Investigaciones Cient\'{\i}ficas (ICMM-CSIC), Sor Juana In\'es de la Cruz 3, 28049 Madrid, Spain.}

\author{Mar Garc{\'i}a-Hern{\'a}ndez}
\affiliation{Instituto de Ciencia de Materiales de Madrid, Consejo Superior de Investigaciones Cient\'{\i}ficas (ICMM-CSIC), Sor Juana In\'es de la Cruz 3, 28049 Madrid, Spain.}

\author{Lin-Lin Wang}
\affiliation{Ames Laboratory and Department of Physics $\&$ Astronomy, Iowa State University, Ames, IA 50011.}

\author{Adam Kaminski}
\affiliation{Ames Laboratory and Department of Physics $\&$ Astronomy, Iowa State University, Ames, IA 50011.}

\author{Paul C. Canfield}
\affiliation{Ames Laboratory and Department of Physics $\&$ Astronomy, Iowa State University, Ames, IA 50011.}

\author{Jos\'e J. Baldov\'i}
\affiliation{Instituto de Ciencia Molecular (ICMol), Universitat de Val\'encia, Catedr\'atico Jos\'e Beltr\'an 2, 46980 Paterna, Spain.}

\author{Isabel Guillam\'on}
\affiliation{Laboratorio de Bajas Temperaturas y Altos Campos Magn\'eticos, Departamento de F\'isica de la Materia Condensada, Instituto Nicol\'as Cabrera and Condensed Matter Physics Center (IFIMAC), Unidad Asociada UAM-CSIC, Universidad Aut\'onoma de Madrid, E-28049 Madrid,
Spain.}

\author{Hermann Suderow}
\affiliation{Laboratorio de Bajas Temperaturas y Altos Campos Magn\'eticos, Departamento de F\'isica de la Materia Condensada, Instituto Nicol\'as Cabrera and Condensed Matter Physics Center (IFIMAC), Unidad Asociada UAM-CSIC, Universidad Aut\'onoma de Madrid, E-28049 Madrid,
Spain.}

\date{\today}

\begin{abstract}
The orthorhombic compound AuSn$_4$ is compositionally similar to the Dirac node arc semimetal PtSn$_4$. AuSn$_4$ is, contrary to PtSn$_4$, superconducting with a critical temperature of $T_c=2.35$ K. Recent measurements present indications for quasi two-dimensional superconducting behavior in AuSn$_4$. Here we present measurements of the superconducting density of states and the band structure of AuSn$_4$ through Scanning Tunneling Microscopy (STM) and Angular Resolved Photoemission Spectroscopy (ARPES). The superconducting gap values in different portions of the Fermi surface are spread around $\Delta_0=0.4$ meV, which is close to but somewhat larger than $\Delta=1.76k_BT_c$ expected from BCS theory. We observe superconducting features in the tunneling conductance at the surface up to temperatures about 20\% larger than bulk $T_c$. The band structure calculated with Density Functional Theory (DFT) follows well the results of ARPES. The crystal structure presents two possible stackings of Sn layers, giving two nearly degenerate polytypes. This makes AuSn$_4$ a rather unique case with a three dimensional electronic band structure but properties ressembling those of low dimensional layered compounds.
\end{abstract}

\maketitle

\begin{figure}
	\begin{center}
	\centering
	\includegraphics[width=1\columnwidth]{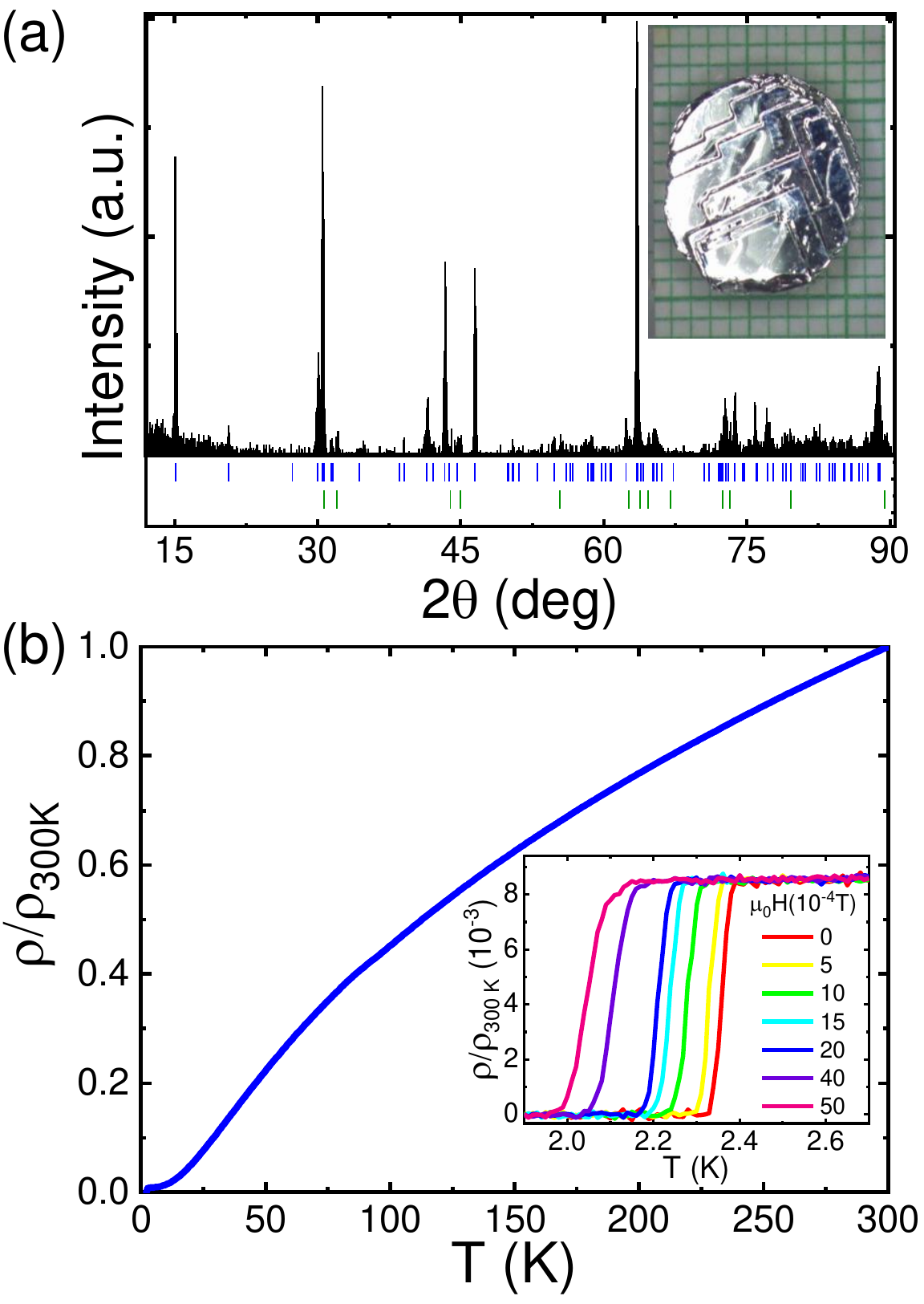}
	\end{center}
	\vskip -0.5 cm
	\caption{In (a) we show the powder X-ray diffraction pattern of AuSn$_4$ as a black line. Vertical blue lines provide the positions expected for the crystal structure of AuSn$_4$. Vertical green lines provide the peaks expected for Sn. A picture of a sample is shown in the inset on a mm sized grid. Notice the flat shape of the crystal. The cleaving plane is parallel to the surface shown in the figure. In (b) we show the resistivity normalized to its room temperature value measured on a four probe configuration. The residual resistance ratio is of 122. The superconducting resistive transition is shown in the inset for different magnetic fields (provided in the legend) applied perpendicular to the surface.}
	\label{FigSamples}
	\end{figure}

\section{Introduction}
Topological insulators host interesting surface states with nontrivial electronic properties, such as dissipationless charge and spin transport along surface channels \cite{RevModPhys.82.3045}. The combination of superconductivity with topological insulators is expected to produce novel behavior, for instance with non-Abelian modes, thanks to the combination of edge conduction channels and the electron-hole symmetry of the superconducting state \cite{RevModPhys.83.1057,doi:10.1146/annurev-conmatphys-030212-184337}. However, superconductivity requires a finite density of states at the Fermi level, which can only be obtained by closing the bulk insulating gap. There is a growing set of semimetallic systems with a finite density of states at the Fermi level that might host nontrivial topological properties, but only a few are superconducting too \cite{doi:10.1146/annurev-matsci-070218-010049,Xiao2021,Burkov2016}.

Often, semimetals with potentially topological crossings in the bandstructure also present a huge magnetoresistance (MR) \cite{Ali2014,Leahy10570,PhysRevB.57.13624,PhysRevB.101.205123}. Among such systems, PtSn$_4$ and PdSn$_4$ stand out because of reports connecting topological properties of the band structure with a three orders of magnitude increase in the resistance between zero field and 14 T \cite{PhysRevB.85.035135,PhysRevB.96.165145,PhysRevMaterials.1.064201}. In addition, the surface of PtSn$_4$ presents Dirac node arcs which consist of a linear dispersion extending over a portion of the Brillouin zone \cite{Wu2016}. PtSn$_4$ and PdSn$_4$ are not superconducting, but AuSn$_4$ is superconducting with a critical temperature of $T_c=2.35$ K \cite{Gendrom1962}. The question if superconductivity in AuSn$_4$ is unconventional is important in the light of the anomalous band structure properties of PtSn$_4$ and PdSn$_4$ \cite{Shen2020,Karn_2022,Sharma_2022}. A recent examination of the bulk and transport properties of single crystals of AuSn$_4$ provides a clear specific heat jump at the superconducting transition. The proposed magnetic field temperature phase diagram provides indications for two-dimensional superconductivity in AuSn$_4$ \cite{Shen2020,Sharma_2022,Karn_2022}.

Here we analyze the normal state and superconducting properties of AuSn$_4$. We grow  single crystals of AuSn$_4$ and characterize them using X-ray diffraction and resistivity. We measure the superconducting properties with scanning tunneling microscopy (STM) and the electronic band structure with angular resolved photoemission (ARPES). The superconducting density of states shows a well opened superconducting gap which remains above the bulk T$_c$. The band structure obtained by ARPES is succesfully compared to Density Functional Theory (DFT) calculations. We discuss the structural properties of AuSn$_4$ and their influence on the observed electronic behavior.

\section{Experiments and methods}

\begin{figure*}
	\begin{center}
	\centering
	\includegraphics[width=1\textwidth]{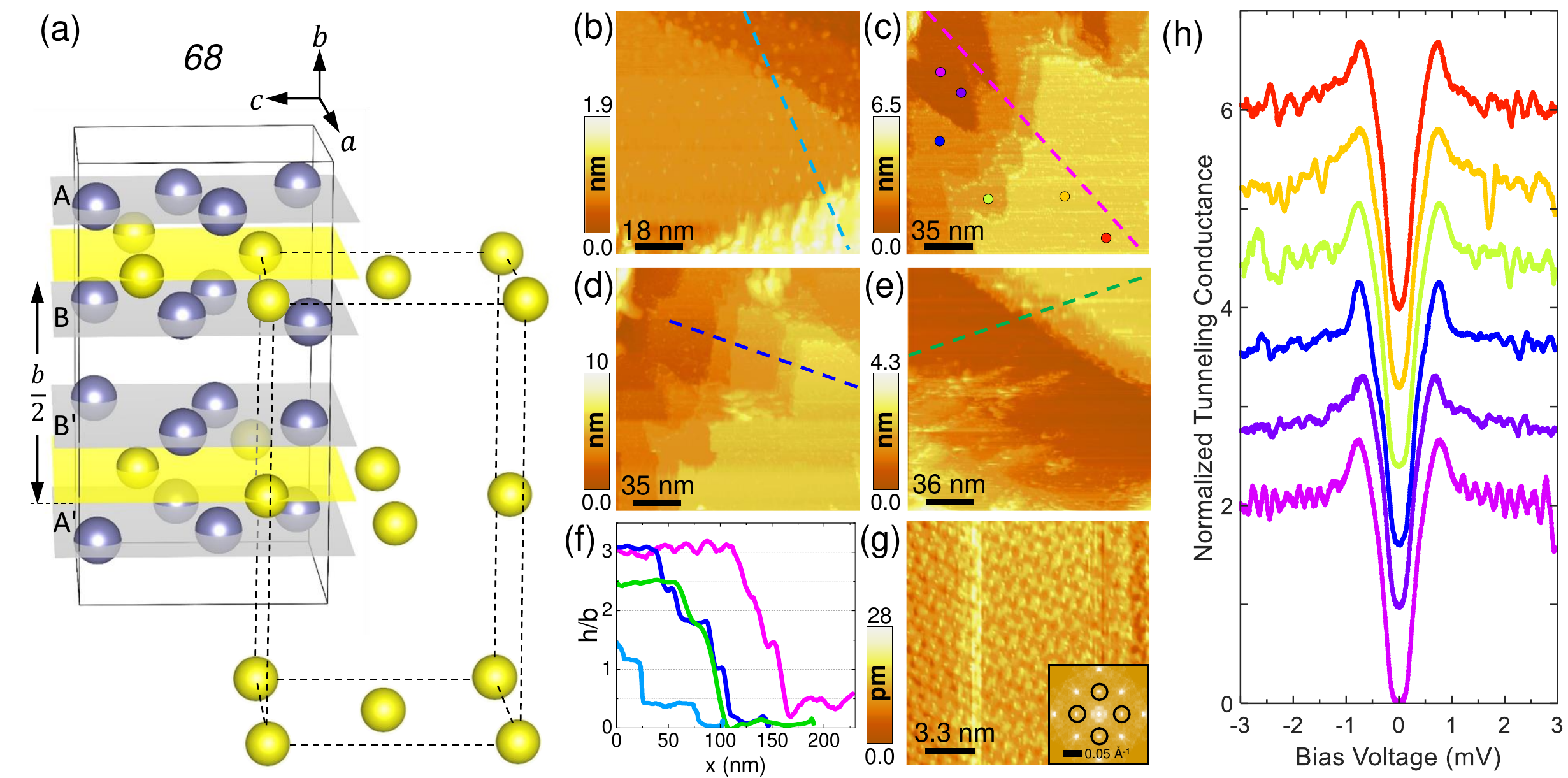}
	\end{center}
	\vskip -0.5 cm
	\caption{In (a) we show the unit cell of AuSn$_4$, corresponding to space group \#68 (black lines). Au and Sn atoms are represented by the gold and grey spheres. The structure of Sn layers, discussed with more detail in the Appendix A, is shown by the A-B-B'-A' sequence. We show the distance between Au (arrow, $b/2$) layers and the Au sublattice (dashed black lines). (b-e) STM images of the AuSn$_4$ surface taken at 100 mK with a bias voltage of 5 mV and a tunneling current of 1 nA. (f) Height vs distance profiles along the colored lines marked on (b-e). The height is normalized to the long axis of the orthorhombic crystal structure (b-axis). (g) Atomic resolution STM image. Its Fourier transform is shown in an inset. Black circles indicate the position of the Bragg peaks corresponding to the Au sublattice. (h) The tunneling conductance is shown as colored lines. Each curve has been taken at different positions of the surface shown in (c) as colored points.}
	\label{FigSTMSurface}
	\end{figure*}

Single crystals of AuSn$_4$ were grown using excess Sn flux \cite{NewMaterialsPhysics}. Based on the binary phase diagram data published in Ref.\,\cite{Okamoto1993}, we mixed high-purity Au (Goodfellow 99.99$\%$) and Sn (Goodfellow 99.995$\%$) with the proportion 88:12 - Sn:Au.  The Au-Sn mixture was introduced in a fritted crucible set \cite{Canfield2016,CanfieldCrucibleSets} and sealed inside a quartz ampoule with Ar gas. In order to fully melt the Au-Sn mixture, the ampoule was heated from room temperature to 1100$^{\circ}$C in 12 h, then cooled to 250$^{\circ}$C in 12 h and finally cooled to 230$^{\circ}$C in 90 h. The ampoule remained for 48 hours at that temperature and was then extracted from the furnace. To separate the remaining liquid from the crystallized AuSn$_4$, we immediately placed the ampoule in a centrifugue \cite{NewMaterialsPhysics}. We obtained plate like crystals with a few mm thickness. As shown in the inset of Fig.\,\ref{FigSamples}(a), the plates can be crucible limited and end up adopting a cryptomorphic (in this case circular/elliptical) morphology that is defined by the inner dimensions of the crucible. Powder X-ray diffraction (Cu K$_{\alpha1},\lambda=1.54 \AA$) on ground AuSn$_4$ crystals led to an orthorhombic crystalline structure with very little difference between two of the three crystalline axis. The observed  $a=0.652$ nm, $b=1.173$ nm, $c=0.652$ nm (Fig.\ref{FigSamples}(a)) were compatible with the crystal structure of AuSn$_4$ found previously. Note that here we find indistinguishable values for $a$ and $c$ from X-ray scattering, although literature reports provide small differences \cite{ZAVALIJ200279,KUBIAK1985339,Kubiak1984,Karn_2022}. Further details about the crystal structure, polytypism and atomic positions are provided in the Appendix A.

To measure the resistivity, we attached four gold wires using silver epoxy to an in-plane shaped sample. In Fig.\,\ref{FigSamples}(b) we show the temperature dependence of the resistivity up to 300 K, from which we obtain a residual resistivity ratio of 122, twice the value found in a recent work \cite{Shen2020}. We estimate a residual resistivity of $\rho_0\sim$ 0.62 $\mu \Omega$ cm, which confirms the excellent quality of the samples. The superconducting transition (inset of Fig.\,\ref{FigSamples}(b)) provides $T_c=2.35$ K in agreement with previous reports \cite{Shen2020,Gendrom1962,Sharma_2022}.

To perform the STM measurements we used the STM and the cryogenic movable sample holder described in Ref.\,\cite{Suderow2011}. We measured the superconducting gap of Al in the same set-up, obtaining a perfect fit to s-wave BCS theory with a measurement temperature of 80 mK and an energy resolution of 8$\mu$V\cite{GUILLAMON2008537,doi:10.1063/5.0059394}. We used a tip of Au, which we cleaned and prepared following Ref.\,\cite{Rodrigo04}. We cleaved the sample below liquid Helium temperature. The sample had a layered shape and was easily cleaved giving shiny flat surfaces oriented perpendicular to the b-axis. The critical magnetic field along this axis (inset of Fig.\,\ref{FigSamples}(b) and Ref.\,\cite{Shen2020}) was of a few hundred Oe. This provides an intervortex distance well above 100 nm at the upper critical field, which is larger than the largest flat surfaces we could find. Thus, we did not follow the tunneling conductance with the magnetic field. Due to the layered structure of the samples, it happened sometimes that pieces of the sample became attached to the Au tip. This gave superconductor-superconductor tunneling conductance curves. In that cases, we de-convoluted the tunneling conductance of the superconducting tip, taking measurements obtained with a normal Au tip, to obtain the tunneling conductance of the sample. To render images and tunneling conductance measurements we used software described in Refs.\,\cite{doi:10.1063/5.0064511,doi:10.1063/1.2432410}.

The ARPES experiments were performed using tunable vacuum ultraviolet (VUV) laser ARPES spectrometer that consists of a Scienta DA30 electron analyzer, picosecond Ti:Sapphire oscillator and fourth-harmonic generator \cite{doi:10.1063/1.4867517}. Data from the laser based ARPES were collected with 6.7 eV photon energy. Angular resolution was set at $\approx$ 0.1$^{\circ}$ and 1$^{\circ}$, along and perpendicular to the direction of the analyzer slit respectively, and the energy resolution was set at 2 meV. The VUV laser beam was set to vertical polarization, i.e. along $k_y$ direction. The diameter of the photon beam on the sample was $\approx$ 15$\mu$m.  Samples were cleaved in-situ at a base pressure lower than 2.2$\times$10$^{-11}$ Torr usually producing very flat, mirror-like surfaces, similar to those obtained in the STM studies. Results were obtained at a temperature of 10.5 K and were reproduced on several different single crystals.

We carried out DFT calculations of the bulk electronic band structure as implemented in the Quantum ESPRESSO package \cite{Giannozzi_2009}. We used the generalized gradient approximation (GGA) and the Perdew–Burke–Ernzerhof (PBE) functional to describe the exchange-correlation energy \cite{PhysRevLett.77.3865}. To fully optimize the structure, we selected standard solid-state US pseudopotentials from the Materials Cloud database \cite{Prandini2018} and sampled the Brillouin zone by a fine $\Gamma$-centered 8 $\times$ 8 $\times$ 8 k-point Monkhorst–Pack mesh \cite{PhysRevB.13.5188}. The ground state and electronic structure of the material were calculated including spin orbit coupling (SOC) by taking fully relativistic norm-conserving pseudopotentials from the Pseudo-Dojo database \cite{VANSETTEN201839}. We sampled the Brillouin zone by a $\Gamma$-centered 10 $\times$ 10 $\times$ 10 k-point Monkhorst–Pack mesh \cite{PhysRevB.13.5188}. The electronic wave functions were expanded with well-converged kinetic energy cut-offs for the wave functions and charge density of 80 Ry and 640 Ry respectively. Dispersion corrections were considered by applying semi-empirical Grimme-D3 corrections \cite{https://doi.org/10.1002/jcc.20495}. The structure was fully optimized using the Broyden-Fletcher-Goldfarb-Shanno (BFGS) algorithm \cite{HEAD1985264} until the forces on each atom were smaller than $1 \cdot  10^{-3}$ Ry/au and the energy difference between two consecutive relaxation steps was less than $1\cdot 10^{-4}$ Ry.

\section{Results}

\subsection{Scanning tunneling microscopy and spectroscopy.}

We show in  Fig.\,\ref{FigSTMSurface}(a) the orthorhombic crystal structure of AuSn$_4$ (space group \#68). In Fig.\,\ref{FigSTMSurface}(b-e) we show several STM images found in different fields of view and in different cool downs. We find flat surfaces separated by steps whose height is an integer or half integer number of the b-axis lattice constant (Fig.\,\ref{FigSTMSurface}(f)). Inside the planes we find atomic size features (Fig.\,\ref{FigSTMSurface}(g)). These can be associated to the Au in-plane sublattice (black circles in Fig.\,\ref{FigSTMSurface}(g)), suggesting that the Au sublattice provides a cleaving plane. This agrees with previous work on PtSn$_4$, which found a Pt sublattice at the surface \cite{https://doi.org/10.1002/anie.201906109}.

\begin{figure}
	\begin{center}
	\centering
	\includegraphics[width=1\columnwidth]{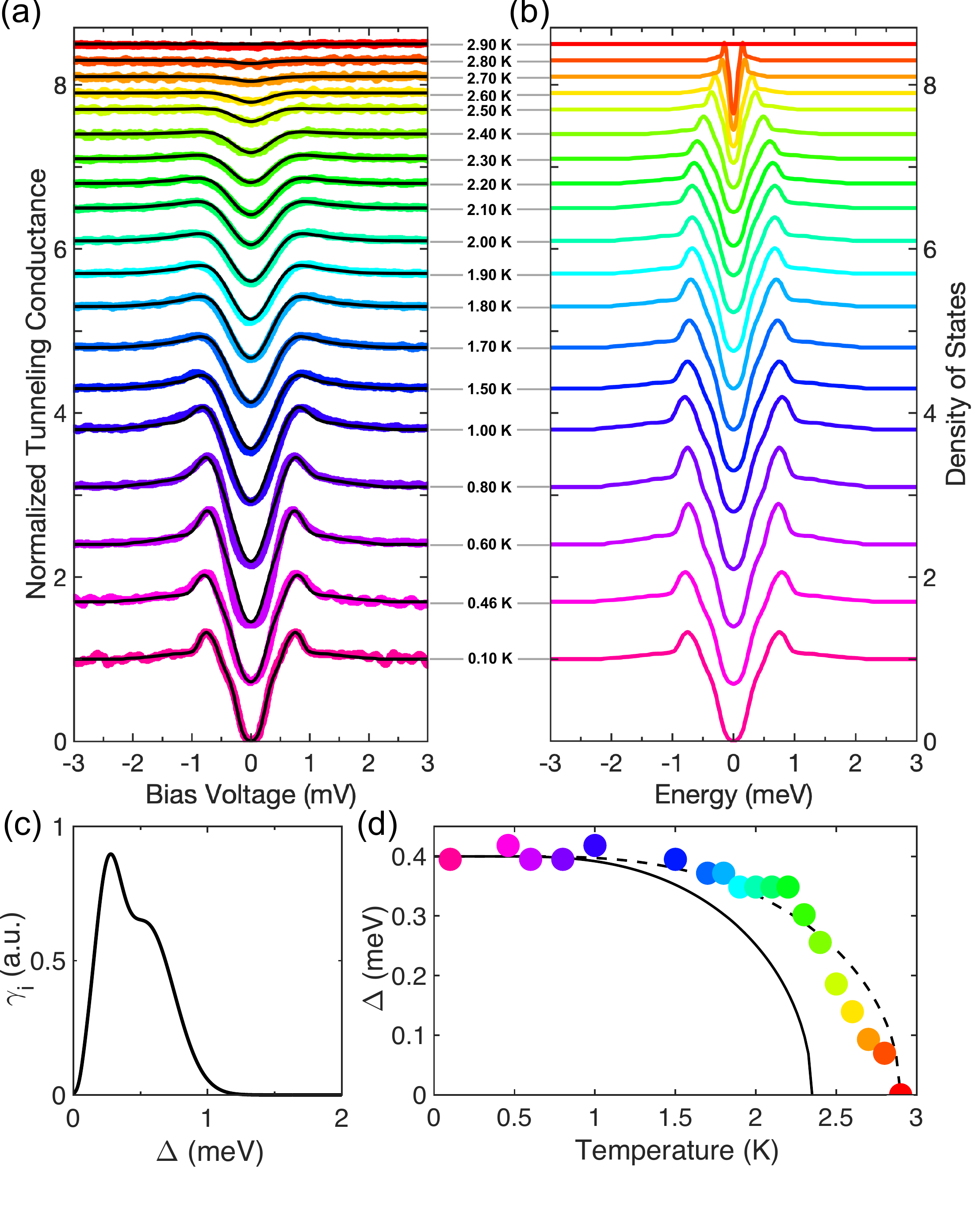}
	\end{center}
	\vskip -0.5 cm
	\caption{In (a) we show tunneling conductance vs bias voltage curves as a function of temperature as colored lines (vertically shifted for clarity). Black lines are the fits obtained from convoluting the density of states vs energy $N(E)$ curves shown in (b) with the derivative of the Fermi function at each temperature. In (c) we show the distribution of values of the superconducting gap which provides (b) as described in the text. In (d) we show the temperature dependence of a weighted average of the superconducting gap. The color of each point in (d) follows the color code of (a,b). The black line in (d) is the expression from s-wave BCS theory using the critical temperature obtained from resistivity $T_c=2.35$ K (continous line) and from STM measurements at the surface (dashed line).}
	\label{FigTunnel}
	\end{figure}

The tunneling conductance vs bias voltage obtained at different positions is shown in Fig.\,\ref{FigSTMSurface}(h). We observe a well opened superconducting gap and pronounced quasiparticle peaks. We show the temperature dependence of the tunneling conductance in Fig.\,\ref{FigTunnel}(a). The superconducting gap value is reduced when increasing temperature, but remains finite above the bulk $T_c$ found in the resistive transition. To obtain the superconducting density of states $N(E)$ as a function of temperature, we deconvolute $N(E)$ from the tunneling conductance $\sigma(V)$, following Refs.\,\cite{Crespo2006,Guillamon08,PhysRevB.97.134501,doi:10.1063/5.0064511,Au2Pb}. $\sigma(V)$ is related to $N(E)$ through $\sigma(V)=\int dE N(E)\frac{\partial f(E-eV)}{\partial E}$, where $E$ is the energy, and $f(E)$ the Fermi function. We seek at each temperature the $N(E)$ that best fits the experiment, showing the corresponding $\sigma(V)$ at each temperature as a black line in Fig.\,\ref{FigTunnel}(a). We show $N(E)$ in Fig.\,\ref{FigTunnel}(b). We find a broad $N(E)$, far from the s-wave single gap BCS expression, $N(E)=\Re(\frac{E}{\sqrt{E^2-\Delta^2}})$. The quasiparticle peaks are broad and there is a large density of states down to energies well below the position of the quasiparticle peaks. 

Note that $N(E)$ has a feature inside the quasiparticle peaks (see Figs.\,\ref{FigSTMSurface}(h) and bottom curves of Fig.\,\ref{FigTunnel}(a,b)). The contribution of this feature changes when taking tunneling conductance curves at different positions (see Fig.\,\ref{FigSTMSurface}(h)). To reproduce $N(E)$ at the lowest temperatures shown in Fig.\,\ref{FigTunnel}(b) we can assume a s-wave BCS density of states built from a distribution of gap values $\Delta_i$, $N(E) \propto \sum_i \gamma_i \Re(\frac{E}{\sqrt{E^2-\Delta_i^2}})$. This provides the distribution $\gamma_i(\Delta_i)$. The resulting $\gamma_i$ vs $\Delta_i$ is shown in Fig.\,\ref{FigTunnel}(c). The distribution consists of two Gaussians, one centered at $\Delta_1=0.54$ meV with a width $\sigma_1=0.21$ meV and another one centered at $\Delta_2=0.25$ meV with a width $\sigma_2=0.1$ meV. Taking bulk $T_c=2.35$ K from resistivity, we find that the BCS value of the superconducting gap $\Delta(T=0K) \approx 1.76 k_B T_{c}$ is of 0.36 meV, in between $\Delta_1$ and $\Delta_2$. Taking $T_{c,surface}=$ 2.75 K, as observed from our tunneling conductance data (curves at the top of Fig.\,\ref{FigTunnel}(a)), we find 0.42 meV, which is also in between $\Delta_1$ and $\Delta_2$. A similar value is found for the position in energy where the density of states is approximately half its normal phase value. We plot the latter as points in Fig.\,\ref{FigTunnel}(d) as a function of temperature, using the curves in Fig.\,\ref{FigTunnel}(b). We compare the result with BCS prediction for the temperature dependence of the superconducting gap, both taking the bulk $T_c$ from the resistivity ($T_c=$2.35 K, continous line in Fig.\,\ref{FigTunnel}(d)) and the temperature of the observed vanishing of the superconducting density of states at the surface ($T_{c,surface}=$ 2.75 K, dashed line in Fig.\,\ref{FigTunnel}(d)).

	\begin{figure}
	\begin{center}
	\centering
	\includegraphics[width=1\columnwidth]{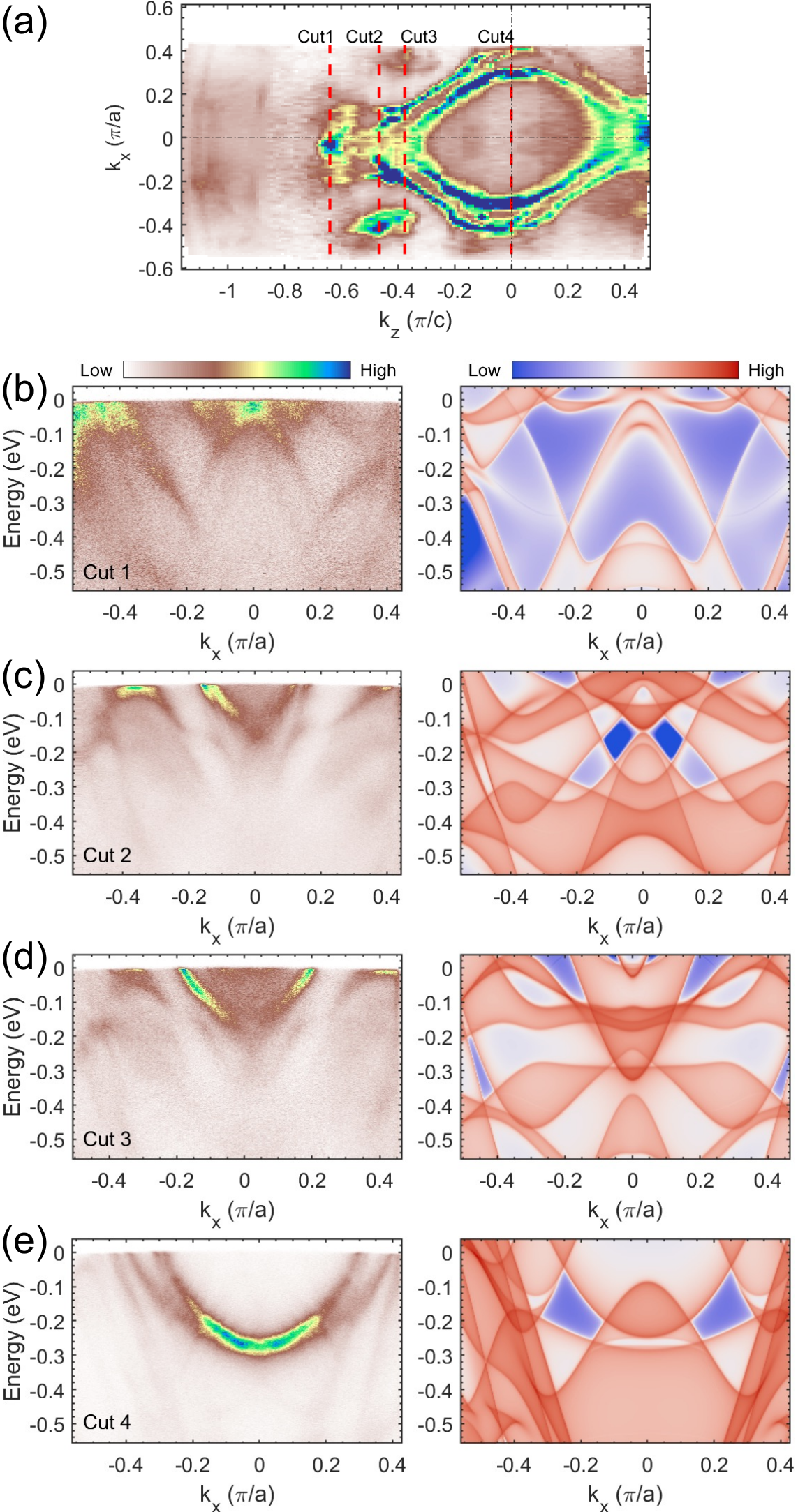}
	\end{center}
	\vskip -0.5 cm
	\caption{(a) ARPES intensity in the plane of AuSn$_4$ close to the Fermi level. The directions of the cuts displayed in subsequent panels are shown by red dashed lines. (b-e) Energy dependence of the electron density along the cuts shown by red lines in (a). Left panels provide the ARPES results and right panels DFT calculations (we represent the bulk band structure projection to the surface plane and use space group \#68, see Appendix B for more details).}
	\label{FigComparison}
	\end{figure}

\subsection{Angular Resolved Photoemission (ARPES) and Density Funcional Theory (DFT) calculations.}

In Fig.\,\ref{FigComparison} we show the photoemission intensity at the Fermi surface (Fig.\,\ref{FigComparison}(a)) and the energy dispersion relation along cuts (red dashed lines in Fig.\,\ref{FigComparison}(a)) made through relevant parts of the Fermi surface (left panels of Fig.\,\ref{FigComparison}(b-e)). We first notice a Fermi surface with a rhombic shape around the center of the Brillouin zone $\Gamma$ (Fig.\,\ref{FigComparison}(a)). We find an intricate band structure. Starting at large $k_z$ (cut 1 in Fig.\,\ref{FigComparison}(a)), we find a small electron like pocket close to $k_x=0$, shown in Fig.\,\ref{FigComparison}(b). There is also a large hole band about 0.1 eV below the Fermi level. When moving $k_z$ towards $\Gamma$, the hole band vanishes and large electron Fermi surfaces take the lead (Fig.\,\ref{FigComparison}(c-e)). A hole pocket is also observed close to $k_z\approx0.6 \pi/c$ and for $k_x\approx0.4 \pi/a$ (Fig.\,\ref{FigComparison}(c-e)).

The main features are mostly found in ARPES as well as in DFT calculations (right panels of Fig.\,\ref{FigComparison}(b-e), further details of the calculations are provided in Appendices B,C). We can identify in the calculations the small electron pocket close to $k_x=0$, with the hole band lying below the Fermi level (Fig.\,\ref{FigComparison}(b)). We also identify how the hole pocket vanishes and the electron like large Fermi surfaces appear when going towards the $\Gamma$ point (Fig.\,\ref{FigComparison}(c-e)). Other aspects of the band structure, such as the band crossing at approximately -0.3 eV shown in Fig.\,\ref{FigComparison}(b), the hole like shape and band crossings observed in Fig.\,\ref{FigComparison}(c,d) close to the Fermi level for $k_x\approx0.3 \pi/a$ or the large electron like shape found close to $k_z\approx0.3 \pi/c$ in the whole energy range (Fig.\,\ref{FigComparison}(e)), show an excellent coincidence between calculations and experiment.

At the $\Gamma$ point (Fig.\,\ref{FigComparison}(e)), DFT calculations show two well separated electron bands. In the experiment, however, we find one electron band which splits with increasing $k_{x,y}$.

\section{Discussion and conclusions}

The observed superconducting density of states suggests a large distribution of gap values over the Fermi surface, going down to values very close to zero energy. Such a large distribution of values of the superconducting gap might come from multiband superconductivity. The obtained intricate shape of the Fermi surface, with many bands crossing the Fermi level (Fig.\,\ref{FigComparison}(b-e)), makes it very likely to have different values of the superconducting gap in different portions of the Fermi surface.

Let us now comment on the superconducting signal in the tunneling conductance we find above $T_c$. Previous work also showed different $T_c$'s in thin films of AuSn$_4$ \cite{KLOKHOLM1966565}. It is useful to compare the crystal structures of AuSn$_4$, PtSn$_4$ and PdSn$_4$. The orthorhombic crystal structure has been determined, finding $a=0.643$ nm, $b=1.135$ nm and $c=0.638$ nm for PtSn$_4$ \cite{TORGERSEN200192}, $a=0.644$ nm, $b=1.145$ nm and $c=0.639$ nm for PdSn$_4$ \cite{https://doi.org/10.1002/chin.200417015}, but $a=0.652$ nm, $b=1.173$ nm and $c=0.652$ nm for AuSn$_4$ \cite{ZAVALIJ200279}. We cannot distinguish the values of $a$ and $c$ from X-ray scattering in AuSn$_4$. We observe a small difference between $a$ and $c$ in the calculated relaxed structure (provided in the Appendix A). Thus, the values of $a$ and $c$ are difficult to distinguish in AuSn$_4$ but can be differentiated clearly in the other two compounds. This suggests that the presence of domains with rotated in-plane orientations is more likely in AuSn$_4$ \cite{KUBIAK1985339,Kubiak1984}. Contrary to crystals of PtSn$_4$ and PdSn$_4$ which present a bulky three dimensional appearance, crystals of AuSn$_4$ have a plate-like shape and are extremely easy to cleave  and very brittle. The Debye temperatures are of 304 K, 255 K and 238 K in PtSn$_4$, PdSn$_4$ and AuSn$_4$ respectively \cite{Vaishnava1989}, suggesting that AuSn$_4$ is softer than PtSn$_4$ and PdSn$_4$. Furthermore, there are significant differences in the thermal expansion in PtSn$_4$ and PdSn$_4$ as compared with AuSn$_4$. Taken together, we can see that the properties of AuSn$_4$ are easily modified by defects along the b-axis and can induce properties characteristic of a layered crystal structure. As we show in the next paragraphs, this is due to polytypic modifications of the structure of AuSn$_4$.

Calculations detailed in the Appendix A show that there is little energy difference between the atomic positions found in two different space groups (\#68 and \#41) in AuSn$_4$. Both present very similar lattice constants and the atomic positions are nearly the same. The Au layers and Au atomic positions are identical. The difference is in the Sn layers above and below the Au layers in Fig.\,\ref{FigSTMSurface}(a). In \#68 Sn layers above and below give an inversion center in the crystal structure (Sn layer stacking sequence A-B-B'-A', with A and A' being inversion symmetric and the same for B and B'), whereas for \#41 the layers above and below have lost the inversion center (Sn layer stacking sequence B-A-B'-A'). The difference between both structures appears within Sn layers. Thus, the two space groups are polytypes of AuSn$_4$. Polytypes form in systems having layers with two structures that are energetically very close. The different arrangements in each layer leads to polytypism. Polytymism can be viewed as a natural sequence of stacking faults which do not produce any other structural changes than a rotation or another symmetry operation on a layer \cite{PhysRevB.69.134111,Palosz+2022+11+34}. The small energy difference between both atomic arrangements in the Sn layers suggests that stacking faults consisting of Sn layers rotated from A to B and viceversa (see Appendix A for more details) are the most probable defects in AuSn$_4$.

The influence of stacking faults in the $T_c$ has been reported in other layered materials. For example, the layered diantimonide LaSb$_2$ shows a very large resistive superconducting transition and the $T_c$ observed at the surface is larger than the temperature at which the resistivity drops to zero \cite{Galvis2013}. Often, transition metal dichalcogenides show no superconductivity at all in certain polytypes, and superconducting properties in others, leading to a spatially dependent superconducting density of states which varies from s-wave BCS theory \cite{D0DT03636F,PhysRevB.82.014518,Guillamon08,Galvis2013,PhysRevB.89.224512,PhysRevB.87.094502,Guillamon2011,PhysRevB.92.134510}. The effect of polytypism on $T_c$ has been studied in the layered dichalcogenide $TaSe_{2-x}Te_x$ \cite{Luo_2015}. In this system, the 3R polytype has shown an increment in $T_c$ up to 17 times the $T_c$ of its 2H polytype; an effect atributed to the change in the electronic properties as a consequence of the small variations in the layer-stacking sequence.

Thus, the presence of intrinsic stacking faults in AuSn$_4$ can explain an increase of $T_c$ by nearly 20\% close to the surface, leading to the observation of superconductivity in the tunneling conductance above the $T_c$ from the resistive transition. Furthermore, we can associate the dependence of the density of states as a function of the position to spatial inhomogeneities of the superconducting properties along the b-axis, perpendicular to the surface.

On the other hand, the coincidence between ARPES results and DFT calculations (presenting the surface projection of the band structure of the \#68 space group, see Fig.\,\ref{Fig41} for the surface projection of the \#41 space group) show that there are only a few aspects in the band structure that are much modified by polytypic stacking faults. The electron bands observed in Fig.\,\ref{FigComparison}(e) are more separated in the DFT calculations than in ARPES. Furthermore, the shape of the Fermi surface is more in-plane isotropic in ARPES (Fig.\,\ref{FigComparison}(e)) than in DFT calculations (shown in Appendix C). But many other features, described above, follow the DFT calculation for the \#68 space group.

We should remind the involved length scales in different measurements. At the photon energy used to collect the ARPES data, the photoelectron escape depth is of about 30 \AA. Therefore the ARPES signal shows the band structure several unit cells close to the surface. Nevertheless, the influence of stacking faults might well extend over a larger scale. The superconducting density of states measured with STM provides an average over distances of order of the coherence length, which is of about 75 nm \cite{Shen2020}. This is large and shows that the polytypic modifications are not just at the last few layers but extend considerably in depth. The magnetoresistance (Appendix D) is also likely influenced by the distribution of polytypic stacking faults over large distances. 

The polytypism of AuSn$_4$ has also consequences that go beyond the pure electronic properties. It has been shown that the layered properties of AuSn$_4$ play a significant role in the embrittlement of Sn based soldering alloys. Indeed, the recent thrust in developing lead-free solders has led to a close re-analysis of phase diagrams of Sn with other elements \cite{doi:10.1063/1.1517165}. In the Au-Sn binary phase diagram, several intermetallic compounds form \cite{CIULIK199371}. AuSn$_4$ is the first compound crystallizing from a Sn rich melt. AuSn$_4$ crystallizes in a melt with a concentration between 7\% and 20\% of Au when crossing the temperature range between 250 $^{\circ}$C and 220 $^{\circ}$C. This easily occurs when soldering wires to electronic appliances because many surface mounting systems finish in a Au containing surface \cite{Chang2006,Ho2000,Ho2002}. The dissolution of Au and its later solidification leads to the precipitation of AuSn$_4$ single crystals, which migrate to the interface between the solder and the substrate. It has been shown that these crystals participate in processes weakening the bond \cite{Ho2000,Ho2002,Lee2004,Alam2004}. The strong tendency to form stacking faults due to polytypism suggests that the pronounced mechanical anisotropy of crystals of AuSn$_4$ could also be related to defects \cite{ZAVALIJ200279,KUBIAK1985339,Kubiak1984}. These might help producing embrittlement at critical parts of the joint \cite{Chang2006,Lee2004}.

In summary, we have grown and characterized single crystals of the superconductor AuSn$_4$. From local Scanning Tunneling conductance measurements we find superconductivity above the bulk $T_c$. ARPES provides an intricate band structure. We discuss the role of polytypism in the electronic and superconducting properties of AuSn$_4$. Polytypism usually appears in quasi two-dimensional superconductors. AuSn$_4$ has however a strongly three-dimensional Fermi surface, suggesting that polytypic stacking faults are relevant to better understand the properties of this system, which is a rather unique example of a three dimensional compound presenting properties characteristic of quasi two-dimensional systems.
\\

\section{Acknowledgments}
We particularly acknowledge very fruitful discussions about the crystal structure of AuSn$_4$ with Enrique Guti\'errez Puebla and Federico Mompe\'an. Work done at Madrid (E.H., B.W., M.A., P.G.T., V.B., C.M., F.M., M.G.G., I.G. and H.S.) was supported by the Spanish State Research Agency (PID2020-114071RB-I00, CEX2018-000805-M), by the Comunidad de Madrid through program NANOMAGCOST-CM (Program No.S2018/NMT-4321), and by EU (PNICTEYES ERC-StG-679080 and COST NANOCOHYBRI CA16218). Segainvex and Sidi at UAM, Madrid, are acknowledged for help in STM construction and sample characterization. Work at Valencia (J.J.B. and A.M.R.) received support from the Plan Gent of Excellence of the Generalitat Valenciana (CDEIGENT/2019/022) and the Excellence Unit “María de Maeztu” CEX2019-000919-M from the Spanish MICINN. Work done at Ames Laboratory (E.O., L.W., A.K. and P.C.C.) was supported by the U.S. Department of Energy, Office of Basic Energy Science, Division of Materials Sciences and Engineering. Ames Laboratory is operated for the U.S. Department of Energy by Iowa State University under Contract No. DE-AC02-07CH11358.

\section*{Appendix}

\subsection{Structure of AuSn$_4$} 

The crystal structure of AuSn$_4$ has been measured using powder X-ray diffraction. All available X-ray scattering data are compatible with both the \#68 centrosymmetric space group $Ccca$ and \#41 noncentrosymmetric space group $Aba2$ \cite{KUBIAK1985339,KUBIAK1981P53,ZAVALIJ200279,Okamoto1993,Sharma_2022,Sharma_2022}. Both present nearly identical lattice parameters, and the difference is purely coming from different heights of different peaks\cite{Karn_2022}. The powder X-ray diffraction is based on milling down the sample into ideally small grains of randomly distributed shapes. The layered properties of AuSn$_4$ make it however quite difficult to mill AuSn$_4$ into powder with random shapes, as it is easy to obtain flakes. This does not lead to an equal distribution of grains, for which the height distribution in powder X-ray diffraction is influenced by the shape of particles in the powder. Thus, it is impossible to distinguish between \#68 and \#41 from powder X-ray diffraction. This might also explain the differences found in the determination of the lattice constants a and c ($a,c\approx 0.65\pm0.01$) in literature\cite{KUBIAK1985339,KUBIAK1981P53,ZAVALIJ200279,Okamoto1993,Sharma_2022,Karn_2022}. As we discuss below, here we show that the ARPES results are better explained by the \#68 space group. Our DFT calculations also show that \#68 is the more stable configuration.

\begin{figure}
	\begin{center}
	\centering
	\includegraphics[width=1\columnwidth]{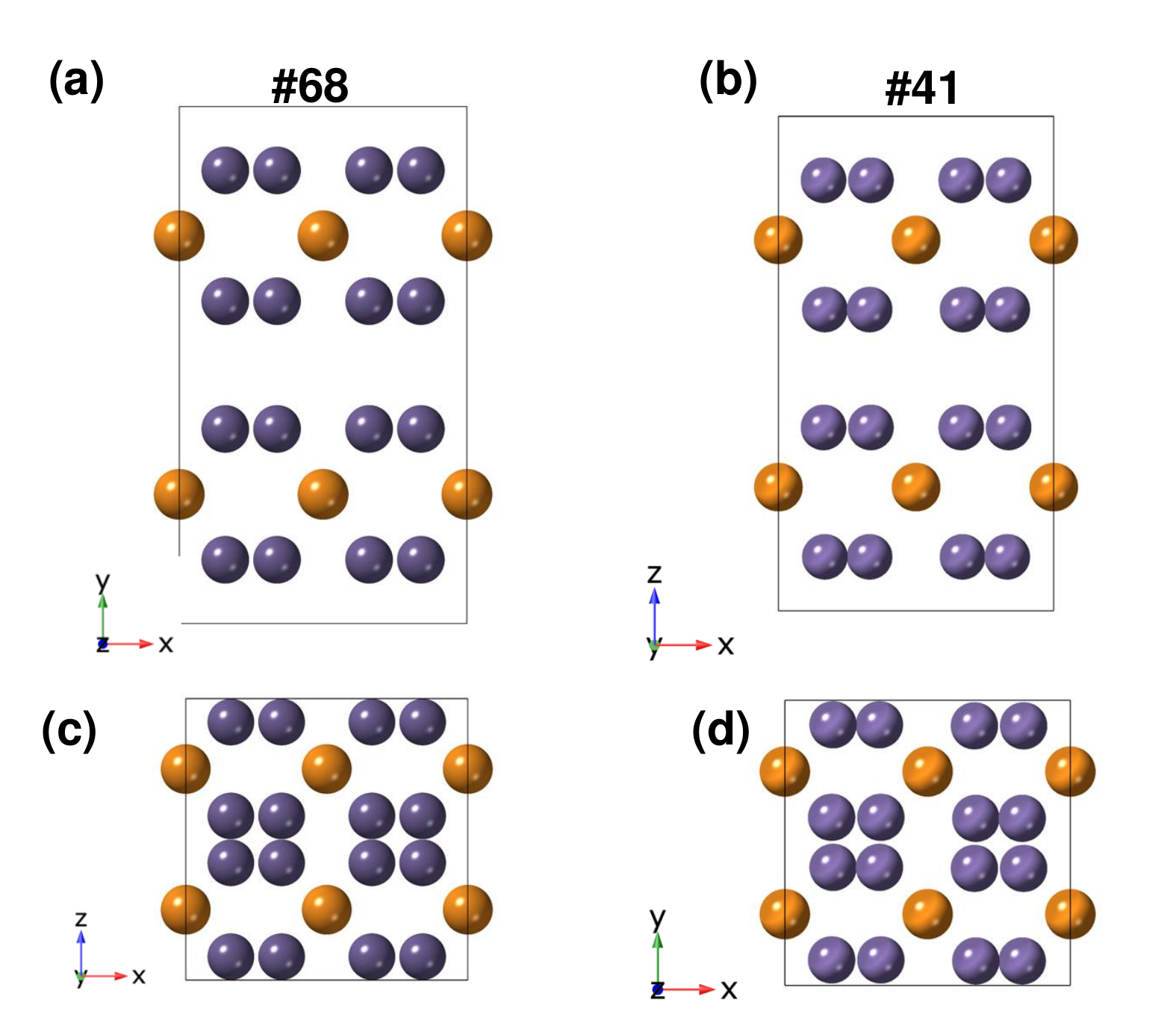}
	\end{center}
	\vskip -0.5 cm
	\caption{We show as grey spheres Sn atoms and Au atoms as yellow spheres. The atomic arrangements are displayed in the same coordinate system. The long axis is defined by the largest lattice parameter, $1.173$ nm. The plane is defined by the two other lattice parameters, both close to $0.652$ nm. As the long axis is the b axis in \#68 and the c axis in \#41, the coordinate systems are chosen as shown in the lower left insets of each panel. In (a) we plot the atomic arrangement of the \#68 space group and in (b) of the \#41 space group, both viewed from the side. In (c,d) we plot the same, but viewed from the top.}
	\label{FigStructures}
	\end{figure}

To describe with more detail the atomic positions in the two possible space groups, we have computed the atomic positions using the same coordinate frame (Fig.\,\ref{FigStructures}). We then identify layers of Au and Sn and describe the arrangement perpendicular to the layers. Here, for simplicity, we focus on the atomic arrangement in Sn layers.

\begin{figure}
	\begin{center}
	\centering
	\includegraphics[width=1\columnwidth]{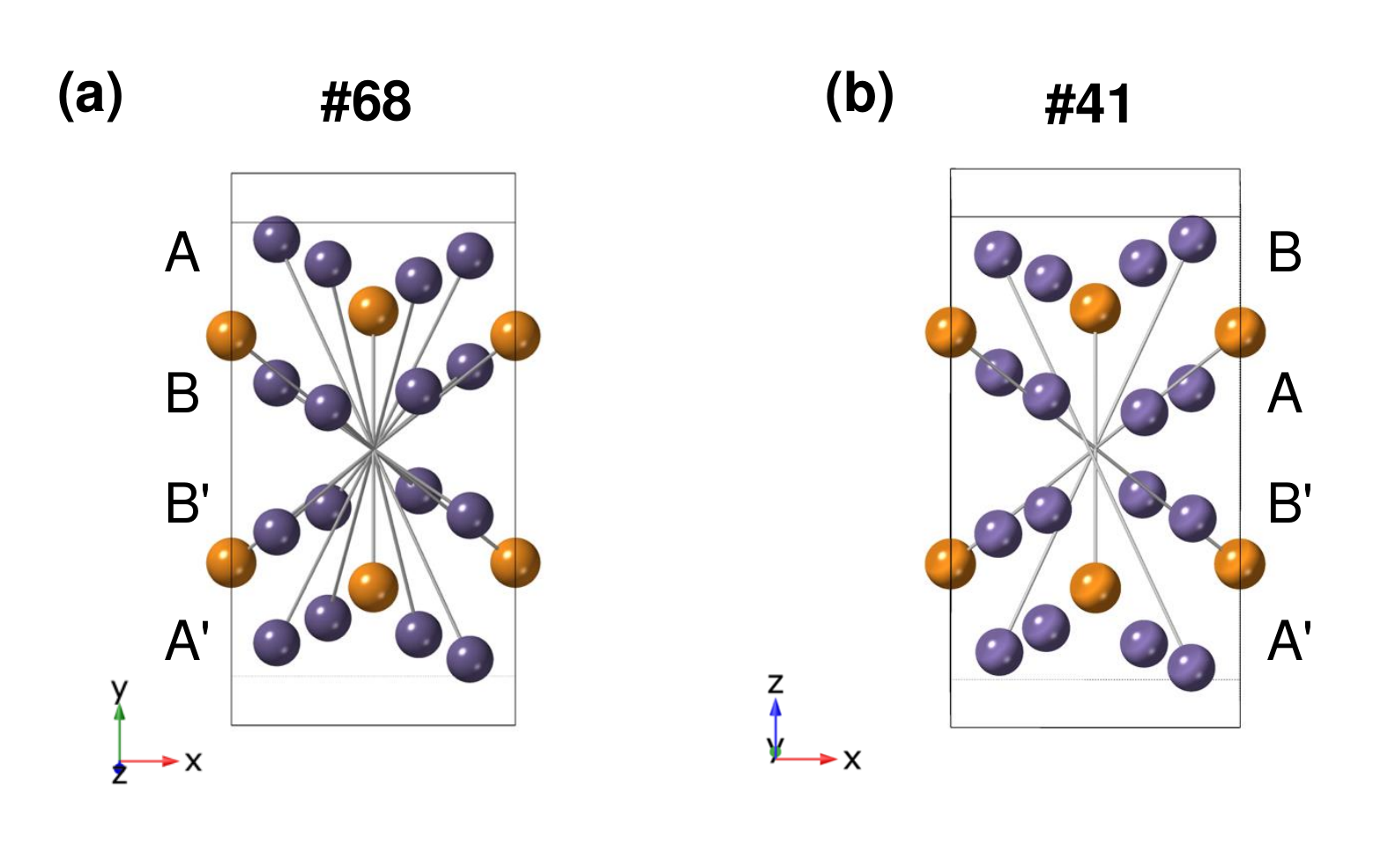}
	\end{center}
	\vskip -0.5 cm
	\caption{We show as grey spheres Sn atoms and Au atoms as yellow spheres. The atomic arrangements are displayed in the same coordinate system, shown in the lower left insets of each panel. Note that there is a tilt with respect to the lateral view shown in Fig.\,\ref{FigStructures}(a,b). Grey lines surrounding the atomic positions provide the orthorhombic shape of the cell displayed in each figure. The other grey tilted lines join atomic positions from between Sn layers in the upper and lower part of the displayed cell. In (a) we plot the \#68 space group and in (b) of the \#41 space group. Note that the latter grey tilted lines cross at the center in \#68 but not in \#41. We also provide the stacking sequence of Sn layers in the view shown in the figure (A-B-B'-A' in \#68 and B-A-B'-A' in \#41).}
	\label{FigStructuresTilt}
	\end{figure}

As we see in Fig.\,\ref{FigStructures}, there is strictly no difference in atomic positions when vieweing both space groups from the side and the top. The atomic positions of Au are indeed identical in both cases. But not the Sn positions. The eight Sn atoms surrounding the Au atoms are arranged differently in each Sn layer. We can see the difference when we slightly tilt the viewpoint (Fig.\,\ref{FigStructuresTilt}). The Sn positions are slightly different in each plane, leading to an inversion center in \#68 and no inversion center in \#41.

\begin{figure}
	\begin{center}
	\centering
	\includegraphics[width=1\columnwidth]{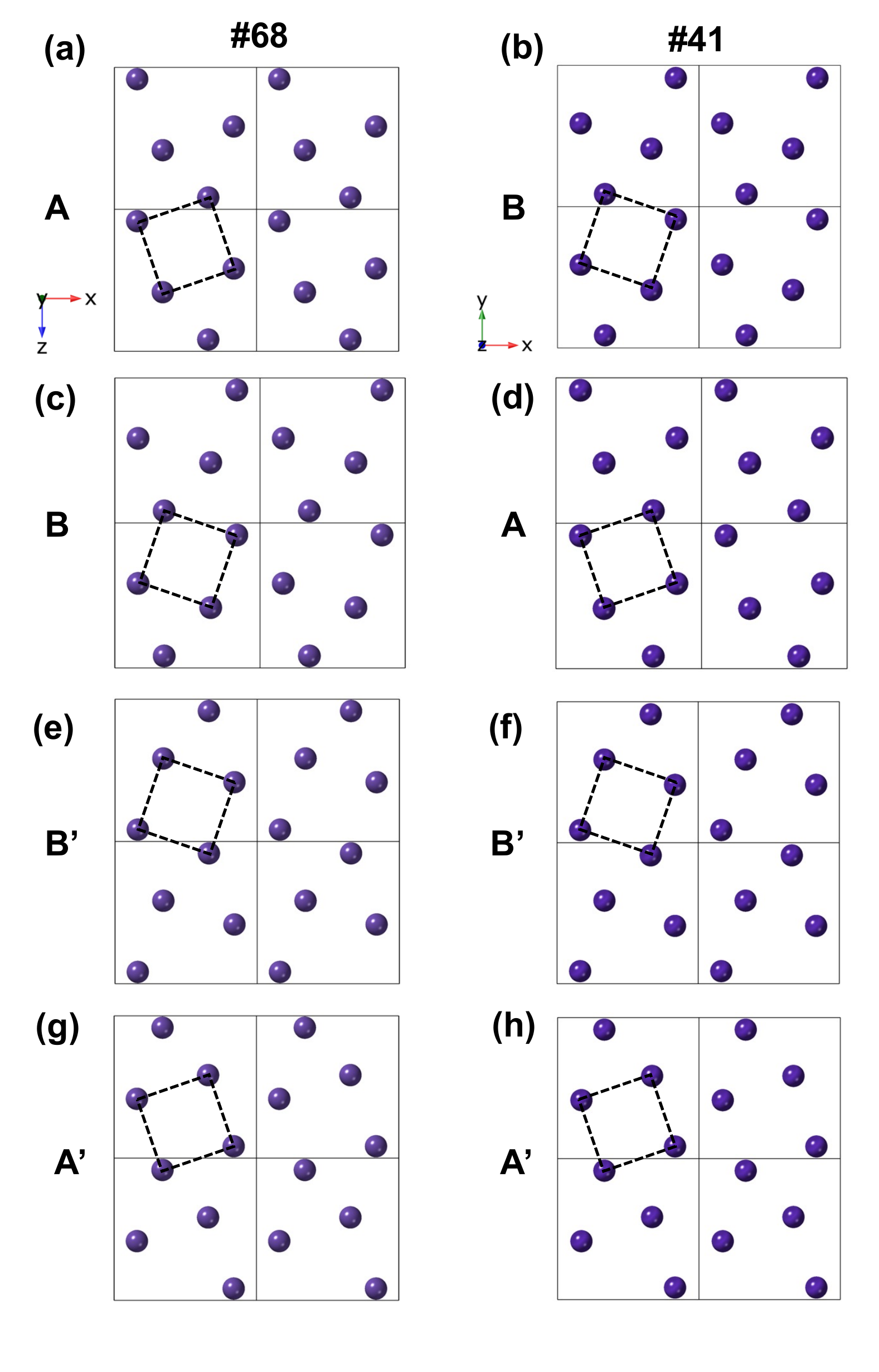}
	\end{center}
	\vskip -0.5 cm
	\caption{We show as grey spheres Sn atoms in each of the four Sn layers shown in previous Figs.\,\ref{FigStructures},\ref{FigStructuresTilt}. Coordinate systems are shown in the lower left insets of (a,b). The four consecutive A-B-B'-A' Sn layers of \#68 space group are shown in (a,c,e,g) and the four consecutive B-A-B'-A' of \#41 space group in (b,d,f,h).}
	\label{FigStackings}
	\end{figure}

A better insight is won when looking at each layer from the top (Fig.\,\ref{FigStackings}). For clarity, we discuss only the Sn positions (Au positions are identical in both space groups, \#68 and \#41). Sn arrangements are made of dumbells of Sn surrounding each Au atom. Within each plane, this leads to squares of Sn that are tilted to the left in A and to the right in B. In A' and B', the tilts are the same for each arrangement, but the Sn squares are displaced. As we see in Fig.\,\ref{FigStackings}(b,d), the stacking sequence is exactly opposite in the upper two layers in \#68 as compared to \#41. This inverted sequence removes the inversion center in the \#41 space group. We provide in Table\,\ref{Table1} the atomic positions of the fully optimized structure.

From this representation of the atomic positions, we see that AuSn$_4$ has two polytypes. The difference between \#68 and \#41 amounts to a different stacking of Sn layers.

\begin{table}[htbp]
\centering
\caption{\label{Table1} Fully optimized atomic coordinates of Au and Sn of the AuSn$_4$ structure in the centrosymmetric space groups Ccca (No 68)(left) and Aba2 (No 41) (right). Atomic positions are scaled with respect the centre of the unit cell now scaled to (0.0, 0.0, 0.0). Numbers in brackets represent indistinguishable points with respect to the inversion center.}
\begin{tabular}{ cc }   
{\bf \# 68} & {\bf \# 41} \\  
\begin{tabular}{ |c|c|c|c| } 
\hline
Atom & x(\AA) & y(\AA) & z(\AA) \\
\hline
\hline
Au(1) &	0.000&	-2.957&	1.640\\
\hline
Au(2)&-3.284&2.957&	1.640\\
\hline
Au(3)&3.284&	2.957&	1.640\\
\hline
Au(1)&0.000&	2.957&	-1.640\\
\hline
Au(3)&-3.284&	-2.957&	-1.640\\
\hline
Au(2)&3.284&	-2.957&	-1.640\\
\hline
Sn(4)&1.063&	4.509&	2.735\\
\hline
Sn(5)&2.221&	-4.509&	2.735\\
\hline
Sn(6)&-2.221&	-1.406&	2.735\\
\hline
Sn(7)&-1.063&	1.406&	2.735\\
\hline
Sn(8)&2.221&	-1.406&	0.546\\
\hline
Sn(9)&1.063&	1.406&	0.546\\
\hline
Sn(10)&-1.063&	4.509&	0.546\\
\hline
Sn(11)&-2.221&	-4.509&	0.546\\
\hline
Sn(8)&-2.221&	1.406&	-0.546\\
\hline
Sn(9)&-1.063&	-1.406&	-0.546\\
\hline
Sn(10)&1.063&	-4.509&	-0.546\\
\hline
Sn(11)&2.221&	4.509&	-0.546\\
\hline
Sn(4)&-1.063&	-4.509&	-2.735\\
\hline
Sn(5)&-2.221&	4.509&	-2.735\\
\hline
Sn(6)&2.221&1.406&	-2.735\\
\hline
Sn(7)&1.063&	-1.406&	-2.735\\
\hline
\end{tabular} &  
\begin{tabular}{ |c|c|c|c| } 
\hline
Atom & x(\AA) & y(\AA) & z(\AA) \\
\hline
\hline
Au(1) &	-3.281&	-1.642&	2.956\\
\hline
Au(2) &	0.000	&1.642&	2.956\\
\hline
Au(3)&	3.281&	-1.642&	2.956\\
\hline
Au(1) &	3.281	&1.642&	-2.956\\
\hline
Au(2) &	0.000&	-1.642	&-2.956\\
\hline
Au(3)&	-3.281&	1.642&	-2.956\\
\hline
Sn&	-2.187&	0.579	&4.411\\
\hline
Sn&	2.187&	2.705	&4.411\\
\hline
Sn&	1.094&	-0.579	&4.411\\
\hline
Sn&	-1.094&	-2.705&	4.411\\
\hline
Sn&	-1.094&	-0.580	&1.310\\
\hline
Sn&	1.094&	-2.705	&1.310\\
\hline
Sn&	2.188&	0.580	&1.310\\
\hline
Sn&	-2.188&	2.705&	1.310\\
\hline
Sn&	-2.187&	-2.705&-1.501\\
\hline
Sn&	2.187&	-0.579	&-1.501\\
\hline
Sn&	1.094&	2.705	&-1.501\\
\hline
Sn&	-1.094&	0.579&	-1.501\\
\hline
Sn&	-1.094&	2.705	&-4.602\\
\hline
Sn&	1.094&	0.58	&-4.602\\
\hline
Sn&	2.188&	-2.705	&-4.602\\
\hline
Sn&	-2.188&	-0.58	&-4.602\\
\hline
\end{tabular} \\
\end{tabular}
\end{table}

The computed difference in energy of atomic positions in both structures is of 0.0165 meV per atom, favoring the \#68 space group. In the relaxed state we find that $a = 0.6567$ nm and $c = 0.6561$ nm in the  \#68 space group and $a = 0.65124$ nm and $b = 0.65680$ nm in the  \#41 space group. 

From such a small difference between \#68 and \#41, we also see that AuSn$_4$ in the \#68 space group is prone to present intrinsic stacking faults consisting of different Sn arrangements within layers.

\subsection{\#41 band structure and surface electronic properties}

\begin{figure}
	\begin{center}
	\centering
	\includegraphics[width=1\columnwidth]{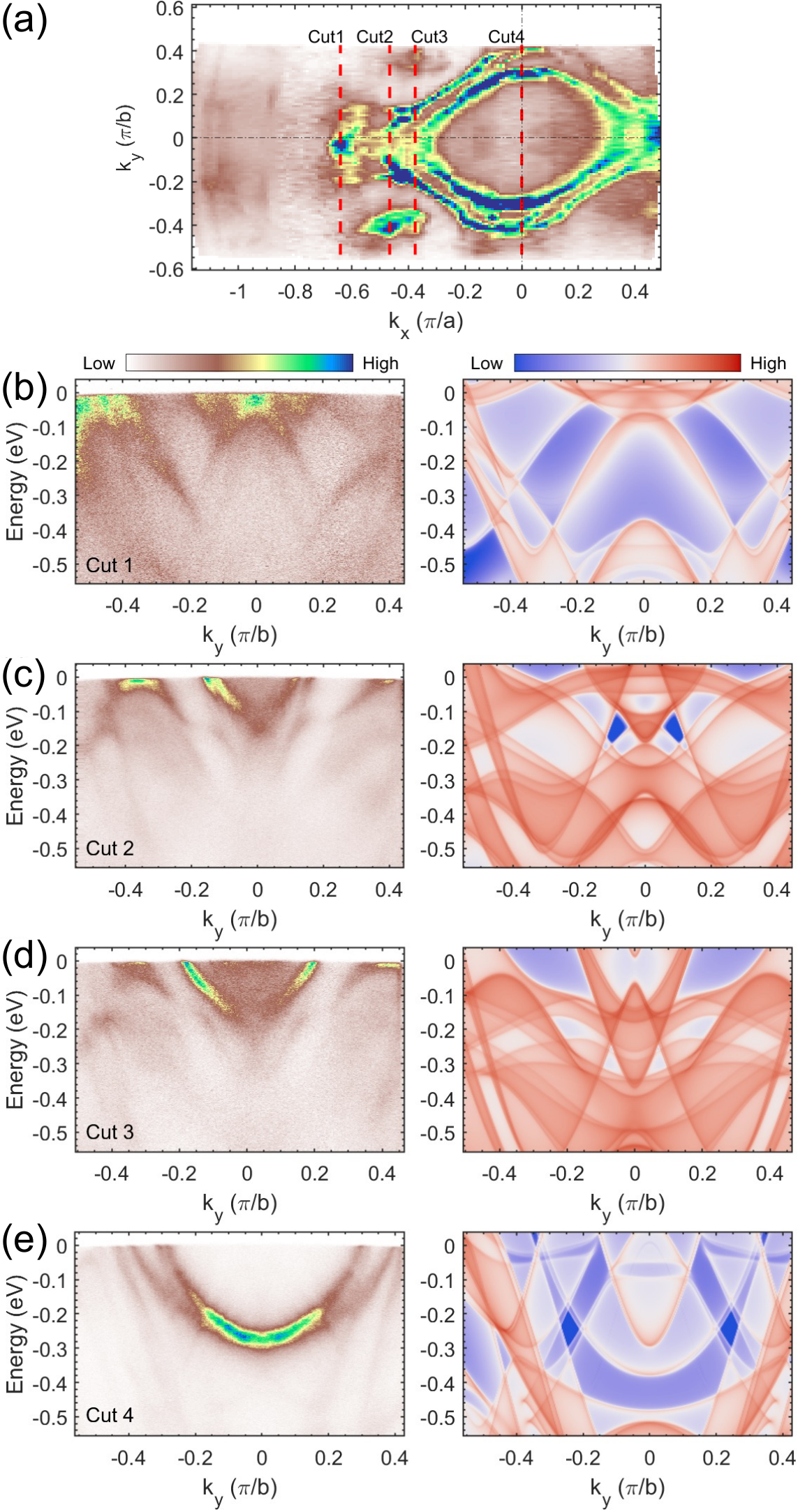}
	\end{center}
	\vskip -0.5 cm
	\caption{(a) AuSn$_4$ bandstructure obtained from ARPES at the Fermi level. (b-e) Comparison between ARPES results and calculations within the \#41 space group (we represent the projection of the band structure to the surface plane, right panels of (b-e)).}
	\label{Fig41}
	\end{figure}

We compare the ARPES results with calculations of the band structure in the \#41 space group in Fig.\,\ref{Fig41}. We see that cut 1 (Fig.\,\ref{Fig41}(b)) reproduces roughly the ARPES results. When approaching the zone center there are significant differences. In particular, the calculated band structure is much more shallow in Fig.\,\ref{Fig41}(d) (cut 3) and the two electron like bands on cut 4 are much more separated  from each other (Fig.\,\ref{Fig41}(e)) than in the \#68 space group band structure discussed in the main text.

\begin{figure}
	\begin{center}
	\centering
	\includegraphics[width=1\columnwidth]{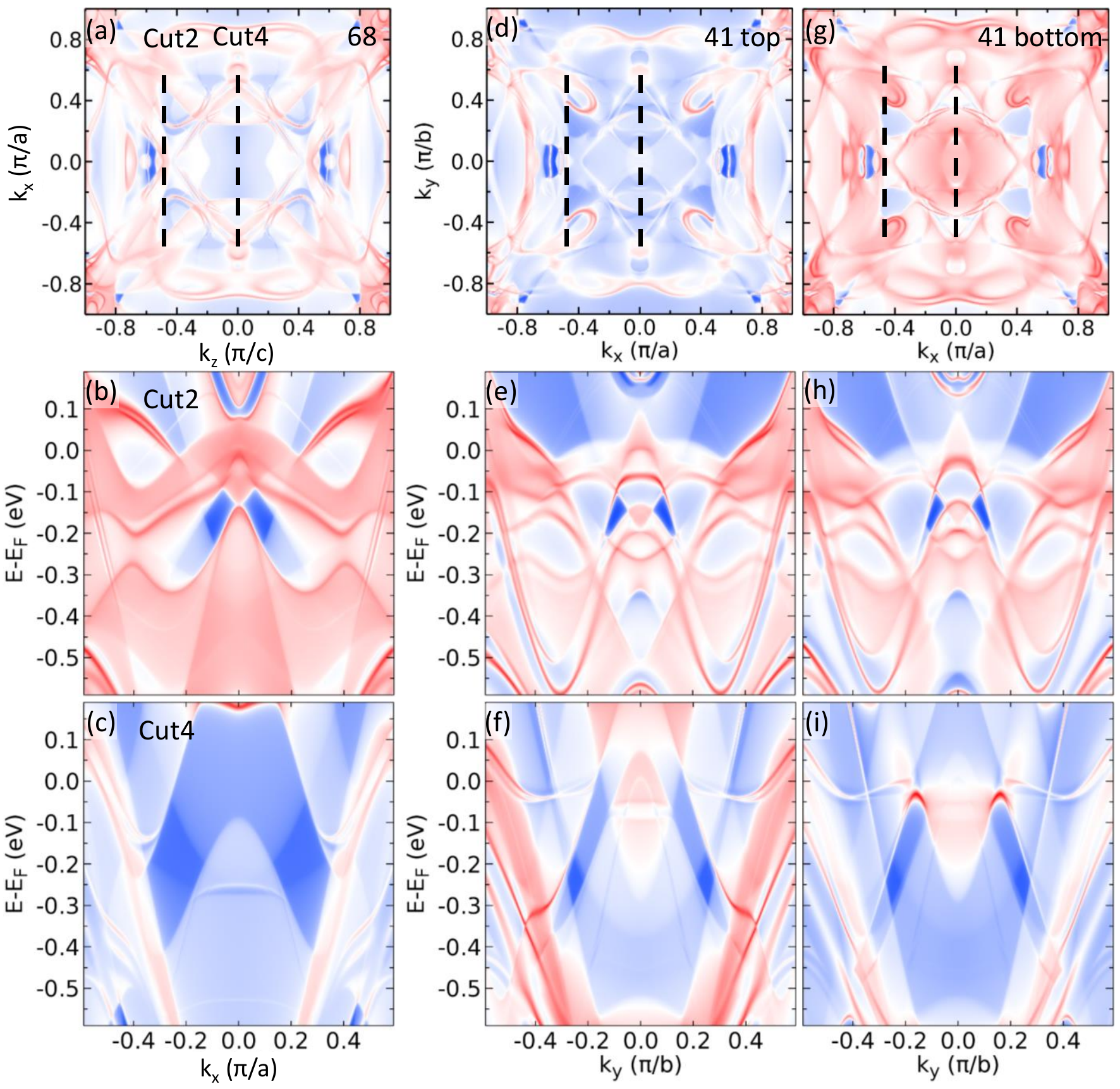}
	\end{center}
	\vskip -0.5 cm
	\caption{(a) Calculated AuSn$_4$ surface band structure at the Fermi level in the \#68 space group. The density of states goes from red (high) to blue (low). The cuts 2 and 4 along the directions shown in Fig\,\ref{Fig41}(a) are shown in (b,c). The surface band structure is shown for the \#41 space group in (d-i). As the \#41 space group is non-centrosymmetric, we show the results for the top (d-f) and bottom (g-i) surfaces. For clarity, we only represent a few cuts, showing that, generally, the comparison with ARPES is worse than the comparison of the bulk band structure.}
	\label{FigDFTSurface}
	\end{figure}

We now discuss the surface band structure. To obtain the surface band structure of AuSn$_4$ in the space groups \#41 and \#68, we constructed maximally localized Wannier functions (MLWFs) \cite{PhysRevB.56.12847, PhysRevB.65.035109} in the Wannier 90 code. We reproduced the bulk band structure in the range $E_F\pm$ 1eV by selecting the Au s,d and Sn s,p orbitals as projectors.  We then calculated the surface spectral functions of semi-infinite surfaces using the surface Green's function method \cite{Sancho_1984,Sancho_1985}, as implemented in Wannier tools \cite{WU2018405}. We show the result of the calculation of the surface band structure for the \#68 space group in Fig.\,\ref{FigDFTSurface}(a-c). For the \#41 space group, the result is in Fig.\,\ref{FigDFTSurface}(d,g). As \#41 is non-centrosymmetric we provide the results of the projection to the two inequivalent top and bottom surfaces. We identify a rhombic shape around the center of the Brillouin zone which is in-plane asymmetric for both structures. Some details of the surrounding features are slightly different. For example, there are pockets around the diagonals of the plane which are more elliptical like in the calculation for the \#41 space group. There are no clear evidences for the two electron like bands near $\Gamma$ observed in the ARPES experiment. The comparison is, in all, much worse.  Note for example the complete absence of a clear set of parabollic bands in cut 4 for all options of the surface band structure. This suggests that ARPES is mostly probing the bulk band structure.

\subsection{Bulk band structure and Fermi surface in AuSn$_4$}

\begin{figure*}
	\begin{center}
	\centering
	\includegraphics[width=1\textwidth]{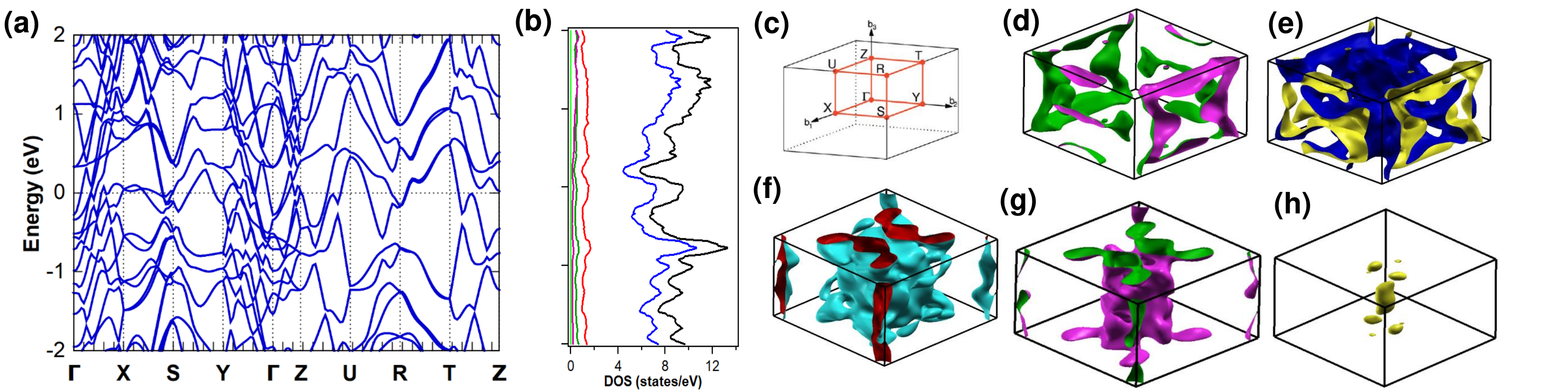}
	\end{center}
	\vskip -0.5 cm
	\caption{In (a) we show the band structure of AuSn$_4$ in the \#68 space group. In (b) we show the k-integrated density of states as a function of the energy in black. The partial densities of states derived from different orbital character is shown as colored lines (blue Sn 5p orbitals, red Sn 5s, dark green Au 5d, violet  Au 6s, light green  Au 5p and grey  Sn 4d, although the latter is in essence zero in the scale of the figure). The Brillouin zone is shown in (c). In (d-h) we show the five Fermi surface sheets we find.}
	\label{FigFS}
	\end{figure*}

The bulk band structure including SOC was calculated using VASP \cite{PhysRevB.54.11169,KRESSE199615} with the lattice constants obtained from X-ray scattering \cite{Kubiak1984,ZAVALIJ200279}. The orbital resolved density of states is shown in Fig.\,\ref{FigFS}(b). We see that the Fermi surface is predominantly of Sn 5p character, with some Sn 5s contribution. The five Fermi surface sheets are shown in Fig.\,\ref{FigFS}(d-h). We see that there is one sheet with intricate pockets at the corners of the Brillouin zone (Fig.\,\ref{FigFS}(d)). Furthermore, there is another large sheet which covers most of the Brillouin zone and has large open Fermi surface sheets  (Fig.\,\ref{FigFS}(e)). Another sheet centered at $\Gamma$ consists of a 3D pocket centered in the Brillouin zone and quasi 2D pockets at the corners of the Brillouin zone (Fig.\,\ref{FigFS}(f)). A similar sheet with slightly smaller pockets is shown in Fig.\,\ref{FigFS}(g). Finally, there are small closed pockets at the center of the Brillouin zone, shown in Fig.\,\ref{FigFS}(h). When projected to the surface, the Fermi surface shows an intricate set of bands. For example, the $\Gamma$ centered pockets in Fig.\,\ref{FigFS}(f,g) provide an in-plane anisotropic shape whose size is similar than the rounded Fermi surface structures observed in ARPES (Fig.\,\ref{FigComparison}(a)). The difference might come from changes in the band structure produced by stacking faults.

Our calculations show that the AuSn$_4$ Fermi surface is different from the Fermi surface obtained in PtSn$_4$. The latter is formed by three small pockets and only one large Fermi surface sheet \cite{Wu2016,PhysRevB.85.035135,Yara2018}, whereas the calculated Fermi surface of AuSn$_4$ is formed by four large and one small sheets. In PtSn$_4$, the surface band structure around the X point presents a sharp band with two Dirac features which are gapped when moving across the X point. This leads to nodes formed by the crossing point of the Dirac dispersion around the X point.

\subsection{Magnetoresistance of AuSn$_4$}

\begin{figure}
	\begin{center}
	\centering
	\includegraphics[width=1\columnwidth]{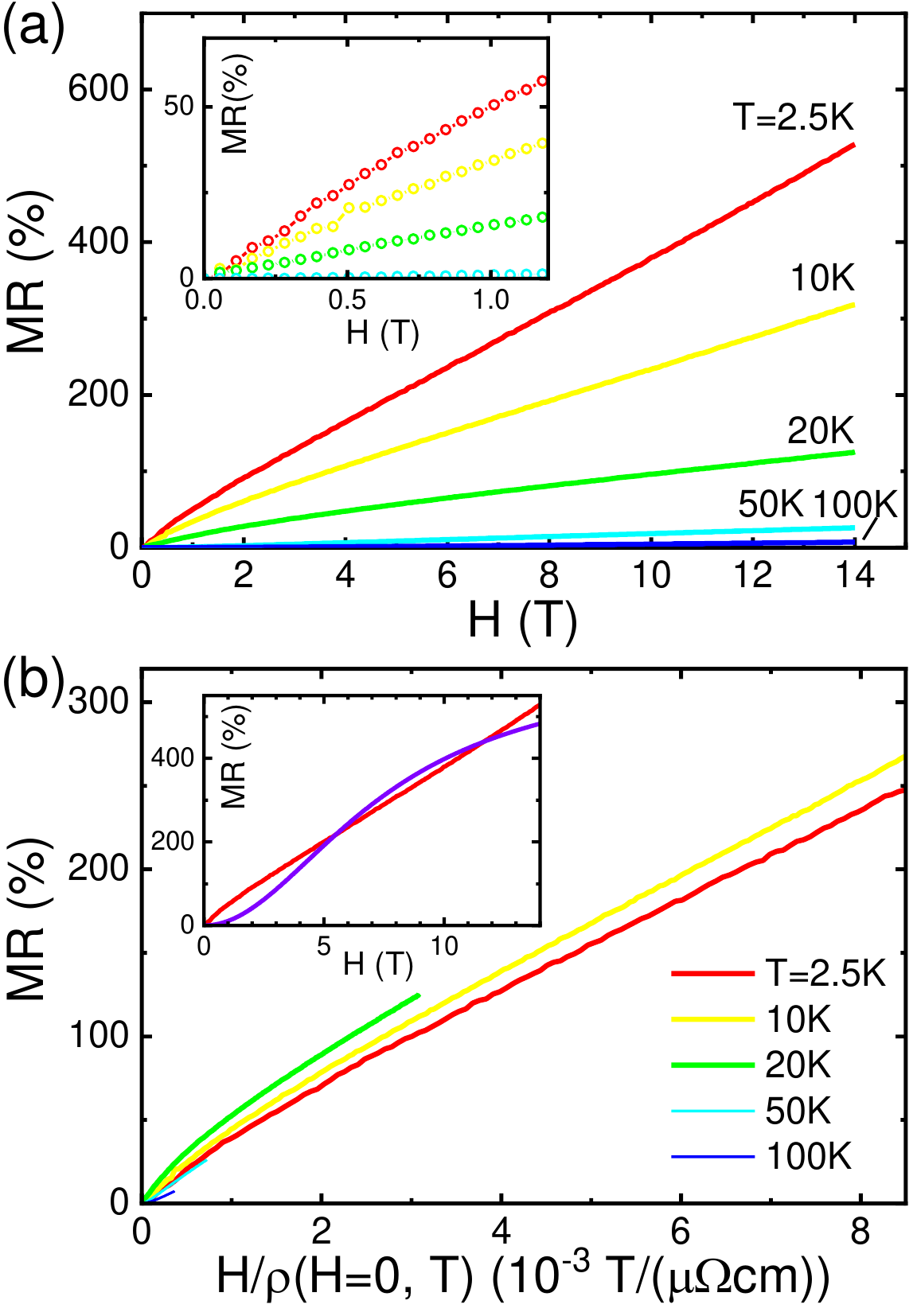}
	\end{center}
	\vskip -0.5 cm
	\caption{In (a) we show the magnetoresistance (MR) of AuSn$_4$ as colored lines for different temperatures. Magnetic field is applied along the $b$ axis. A zoom for low magnetic fields is shown in the inset. In (b) we show the same data in a Kohler's plot, i.e. plotted as a function of the magnetic field normalized to the zero field temperature dependent resistivity $\rho(H=0,T)$. The inset shows the MR at low magnetic fields and at 2.5 K (red line) together with a fit to a simple two band model discussed in the text (purple line).}
	\label{FigMR}
	\end{figure}

In Fig.\,\ref{FigMR}(a) we show the magnetoresistance (MR) $\left( MR=\frac{\Delta\rho(H,T)}{\rho(H=0,T)}\right)$ as a function of the magnetic field up to 14 T at different temperatures between $2.5$ K and $100$ K. The MR increases by more than 500\% between 0 T and 14 T at 2.5 K. In the inset of Fig.\,\ref{FigMR}(a) we provide the MR at small magnetic fields.

The MR changes as a function of temperature. At temperatures below 20 K we observe a slight downward curvature below about 2 T. Above 2 T, we obseve a linear increase up to 14 T without any sign for saturation. Above 20 K we observe that the MR is significantly reduced and the magnetic field dependence approaches the quadratic behavior observed in simple metals for the whole field range. We provide Kohler's plot in Figure \ref{FigMR}(b). This compares the temperature dependence of the resisivity $\rho(H=0,T)$ with the MR, by normalizing the magnetic field to $\rho(H=0,T)$. We see that Kohler's scaling roughly holds, as curves at all temperatures are quite close to each other. Nevertheless, we also observe that the downward curvature below 20 K and the quadratic behavior at high temperatures produce visible deviations from Kohler's rule.

The MR in AuSn$_4$ is large, but considerably smaller than the MR of PtSn$_4$ and PdSn$_4$ \cite{PhysRevB.85.035135,PhysRevB.96.165145,PhysRevMaterials.1.064201}. Other than a slight downward curvature below 2T, AuSn$_4$ shows a linear and nonsaturating behavior of the MR, in contrast to a quadratic-like MR observed in PtSn$_4$ and PdSn$_4$. To understand the MR, we can try to use a simple two-band model with near electron-hole compensation, succesfully used previously in PtSn$_4$ and PdSn$_4$, see Refs.\,\cite{PhysRevB.85.035135,PhysRevB.96.165145,PhysRevMaterials.1.064201}.

The two-band model assumes that there is no anisotropy in the Fermi surface, and that all electron or hole-like carriers have the same mobility. Within this model, the MR would show a close to a $B^2$ behavior or be saturated at high field depending on the level of the electron/hole compensation. Neither of the two cases can explain the non-saturating linear MR observed here. The inset of Fig.\,\ref{FigMR}(b) shows as a purple line the best fit that can be obtained from the two-band model, manifesting a large deviation from the experimental data (we use $\rho(B)=\frac{n_e\mu_e+n_h\mu_h+\mu_e\mu_h(n_e\mu_h+n_h\mu_e)B^2}
{e[(n_e\mu_e+n_h\mu_h)^2+(n_e-n_h)^2\mu_e^2\mu_h^2B^2]}$
where $n_e$ ($n_h$) and $\mu_e$ ($\mu_h$) are electron (hole) concentration and mobility \cite{Pippard1989,Ali2014}). Our calculations show that the Fermi surface of AuSn$_4$ contains five large and highly anisotropic sheets (Fig.\,\ref{FigFS}).  Moreover, four out of five sheets are open (Fig.\,\ref{FigFS}(d), (e), (f), and (g)). This suggests that the two band model may not be sufficient to explain the MR. 

We can also consider other sources for a nonsaturating MR, such as a linearly dispersing portion of the bandstructure, charge density waves, open orbits or disorder  \cite{Pippard1989,Ali2014,Abrikosov1998,Abrikosov2000,Assaf2013,Eto2010,Feng2019,Gibson2014,Hikami1980,Kisslinger2015,Gopinadhan2015,Liang2014,Parish2003,Narayanan2015,Shekhar2015,PhysRevResearch.2.022042,PhysRevB.101.205123,FernandezLomana2021,Leahy10570,PhysRevB.57.13624,PhysRevB.101.205123}. We do not find portions with a linear dispersion close to the Fermi level in the band structure. A charge density wave would lead to a feature in the temperature dependence of the resistivity, which we do not observe. Open orbits are clearly present in the large and complex multiband Fermi surface, but the mixture with other bands make it unlikely that these play a significant role in the MR. The remaining mechanism is disorder, here most probably related to the polytypic nature of the crystal structure. Notice that we do not observe the dip found at low magnetic fields in Refs.\,\cite{Shen2020,Sharma_2022}, but instead a smooth increase of the MR with the magnetic field. This suggest that different sets of samples have a different density of stacking faults, indicating that polytypism induced disorder has an influence on the MR.

\bibliographystyle{apsrev4-1-titles}

%

\end{document}